\documentclass[aps,preprint]{revtex4}

\usepackage{graphicx}% Include figure files

\begin{document}
\title{Control of cardiac alternans in a mapping model with memory}
\author{Elena Tolkacheva$^{1}$, M\'onica M. Romeo$^{2}$, Daniel J. Gauthier$^{1,3}$}
\affiliation{$^{1}$Department of Physics, $^{2}$Mathematics Department, $^{3}$Department
of Biomedical Engineering, and Center for Nonlinear and Complex Systems, Duke University,
Durham, North
Carolina 27708\bigskip , USA}
\date{\today}

\begin{abstract}
We investigate a map-based model of paced cardiac muscle in the presence of
closed-loop feedback control. The model relates the duration of an action
potential to the preceding diastolic interval as well as the preceding
action potential duration and thus has some amount of `memory.' We find that
the domain of control depends on this memory, independently of the specific
functional form of the map. The memory-dependent domain of control can
encompass large feedback gains, thus providing the first possible
explanation of the recent experimental results of Hall {\it et al.} [Phys.
Rev. Lett, {\bf 88}, 198102 (2002)] on controlling alternans in small pieces
of rapidly-paced cardiac muscle.
\end{abstract}

\maketitle
Sudden cardiac death, primarily caused by ventricular arrhythmias, is a
major public health problem: it is one of the leading causes of mortality in
the western world. A possible precursor of some arrhythmias is the
beat-to-beat variation of the cardiac electrical excitation occurring at
fast heart rates \cite{karma,gilmour1}. This variation appears as a sequence
of long-short-long-short cycles (termed alternans) in important
physiological characteristics such as action potential duration (APD) and
conduction time. A focus of recent theoretical and experimental studies is
to investigate the mechanisms causing alternans and to terminate this
response pattern using closed-loop feedback methods developed by the
nonlinear dynamics community. Suppressing such alternans may then prevent
the onset of fibrillation~\cite{karma1}.

Over the last few years, several studies have demonstrated that alternans
can be suppressed with dynamic feedback control of the pacing interval \cite
{dan,hall,karma2,dan1,christini}. Control of alternans in the conduction
time across the atrioventricular (AV) node has been demonstrated in both 
{\it in vitro} rabbit hearts \cite{hall} and {\it in vivo} human hearts \cite
{christini}. The observed AV-nodal alternans are known to be well described
by a one-dimensional map-based mathematical model \cite{christini}. The
model can be used to predict the range of control parameters that stabilize
the desired response patterns.

Recently, Hall {\it et al.} \cite{dan} demonstrated successful control of
alternans in small pieces of {\it in vitro} paced bullfrog ventricles.
Understanding how to control ventricular myocardium is important because it
is the primary substrate for fibrillation. Controlling alternans in
ventricular myocardium was expected to be more difficult than controlling
AV-nodal alternans because past research suggests that its dynamics is more
complicated. On the contrary, the experiments described in Ref. \cite{dan}
demonstrated that alternans could be suppressed over a wide range of control
parameters and over the entire range of pacing rates for which alternans was
observed. The observations were compared to the predictions of two map-based
mathematical models. Fitting the bifurcation diagrams of the mathematical
models to the experimental data did not produce good fits for their domains
of control. Specifically, control of alternans was observed for feedback
gains as large as four in the experiments, whereas the models predicted that
the gain must be less than $\sim 0.4$ and limited to a small region of
pacing rates near the bifurcation to alternans.

The primary purpose of this Letter is to analyze a map-based model of paced
cardiac muscle in the presence of closed-loop feedback control. The model
contains some degree of `memory' so its dynamics displays higher-dimensional
behavior. We find that the domain of control can encompass large feedback
gains, qualitatively consistent with the experiments of Hall {\it et al.}
Our results demonstrate that higher-dimensional effects can enhance the
effectiveness of control under the appropriate conditions. This may lead
eventually to the development of methods for {\it in vivo} control of
whole-heart functions. Future experiments are needed to determine whether
the specific form of memory considered here is displayed by cardiac tissue.
Nevertheless, our results indicate the importance of memory in the control
of alternans.

To set the stage for understanding the behavior of the higher-dimensional
model, we first consider a simpler one-dimensional map that contains no
memory. This mapping model is given by \cite{nolasco,guevara} 
\begin{equation}
A_{n+1}=f(D_{n}),  \label{eq:oldmap}
\end{equation}
where $A_{n+1}$ is the APD generated by the $(n+1)^{st}$ stimulus, and $%
D_{n} $ is the $n^{th}$ diastolic interval (DI), {\em i.e.} the interval
during which the tissue recovers to its resting state after the end of the
previous ($n^{th}$) action potential. We illustrate a typical bifurcation
diagram in Fig. 1a. Note that the tissue has a stable 1:1 response pattern
(every stimulus elicits an action potential of equal duration) for long
pacing intervals (slow pacing rate). As the pacing interval decreases
(faster pacing), the 1:1 response becomes unstable and a transition to
alternans (2:2 response) occurs. At even faster pacing rates, the 2:2
response pattern becomes unstable and is replaced by a stable 2:1 response
where an action potential is produced by every second stimulus. For the 1:1
and 2:2 responses considered herein, the pacing relation between APD and DI
is $D_n=B-A_n$, where $B$ is the basic pacing interval.

For this simple model, the transition from 1:1 to 2:2 behavior shown in the
Fig. 1a can be understood by investigating the restitution properties of the
cardiac membrane. Specifically, to predict the pacing rates at which the 1:1
response is stable, one constructs the restitution curve (RC) by plotting
APD as a function $f$ of the preceding DI, as in (\ref{eq:oldmap}). Nolasco
and Dahlen \cite{nolasco} proposed that the transition to alternans occurs
when the slope of the RC is equal to unity. However, recent studies \cite
{dan,dan2,fox,gray,lena1} have shown that the slope of the RC at the onset
of alternans can be significantly larger than unity and thus this criterion
fails to predict the existence of alternans in some cases. Later in the
paper, we analyze the map (\ref{eq:oldmap}), independently of the specific
form of the function $f$, in the presence of closed-loop feedback control
and discuss why it fails to explain the experimental observations.
\begin{figure}
\includegraphics{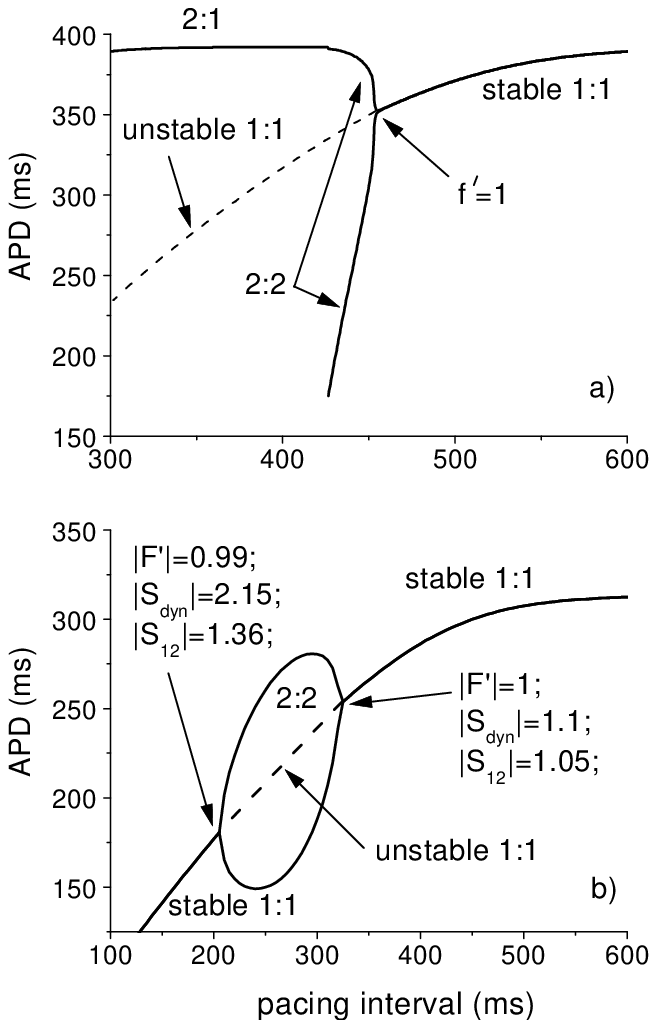}% Here is how to import EPS art
\caption{\label{fig:epsart} 
Bifurcation diagrams showing existence of alternans for a) the
memoryless mapping model (1) and b) the mapping model (2) with memory.
Arrows indicate the points where slopes are determined. Dashed lines show
unstable 1:1 response. For a), $f(D_n)=A_1-A_2\exp (-D_n/\protect\tau )$,
where $A_1=392$ ms, $A_2=525.3$ ms, $\protect\tau =40$ ms [9]. For the
mapping model (2), a specific form of the function $F$ is taken from Refs.
[14, 18], where all parameters have their typical values except $\protect\tau%
_{sopen}=50$ ms.}
\end{figure}
In contrast to the memoryless model (\ref{eq:oldmap}), we find that a
cardiac mapping model of the form 
\begin{equation}
A_{n+1}=F(A_{n},D_{n})  \label{eq:map}
\end{equation}
can possess a domain of control that encompasses large control gains,
independently of the specific form of $F$. This model has some
amount of memory because it relates the duration of the next action
potential both to the previous DI and to the previous APD, {\it i.e.}, the
history of the dynamical system. Several previous studies \cite
{dan2,otani,chialvo,gilmour2} indicate that higher-dimensional behavior
(so-called memory effects) is present in real cardiac tissue and thus has to
be taken into account in order to correctly predict the onset of alternans.
The general form of this model was first introduced by Otani and Gilmour 
\cite{otani} to explain empirical observations from paced dog cardiac
Purkinje fibers. More recently, a specific form of $F$ was derived
analytically \cite{lena} from a three-current ionic model \cite{FK}.

Our analysis of the model (\ref{eq:map}) shows that it displays
rate-dependent restitution so that there exist two primary types of RCs (the
dynamic and S1-S2 RCs), which can be measured independently using different
pacing protocols \cite{lena1}. In the {\em dynamic pacing protocol}, the
pacing interval $B_{1}$ is held fixed until the tissue reaches equilibrium,
and then progressively shortened. This yields pairs of steady-state values ($%
A^{\ast },D^{\ast }$) for each $B_{1}$. In the {\em S1-S2 pacing protocol,}
a premature stimulus (``S2'') is delivered at an interval $B_{2}$ after
pacing the tissue with a sufficiently large number of ``S1'' stimuli at a
pacing interval $B_{1}$ so that the tissue reaches equilibrium. \ The S1-S2
RC is determined by measuring the resulting APD for various coupling
intervals $B_{2}$. Experimental studies have shown that the S1-S2 and
dynamic RCs differ significantly, and have different slopes (denoted as $%
S_{12}$ and $S_{dyn},$ respectively). This is consistent with the
predictions of the mapping model (\ref{eq:map}). The transition to alternans
is governed by the combination of the slopes $S_{dyn}$ and $S_{12}$, so that
alternans can exist when 
\begin{equation}
\left| F^{\prime }\right| =\left| 1-\left( 1+\frac{1}{S_{dyn}}\right)
S_{12}\right| \geq 1,  \label{eq:criterion}
\end{equation}
where $F^{\prime }$ is the full derivative of $F$, with respect to $A_{n}$,
evaluated at a fixed point \cite{lena1}. Note that $S_{dyn}=S_{12}$ when
there is no memory in the model, and that they differ substantially when
there is large memory. A typical bifurcation diagram showing alternans in
the mapping model (\ref{eq:map}) is presented in Fig. 1b, where the slopes
of the RCs are different and do not equal unity at the onset of alternans.

We now consider control of alternans in both models (\ref{eq:oldmap}) and (%
\ref{eq:map}). To suppress alternans and stabilize the 1:1 pattern, Hall 
{\it et al.} \cite{dan} adjusted the pacing period by an amount given by 
\begin{equation}
\varepsilon _{n}=-\gamma (A_{n-1}-A_{n-2})  \label{eq:epsilon}
\end{equation}
where $\gamma $ is the feedback gain. Control is initiated by adjusting the
basic pacing interval $B$ by $\varepsilon _{n}$. Applying the control
technique to the mapping model (\ref{eq:oldmap}) yields 
\begin{equation}
A_{n+1}=f(D_{n})=f(B+\varepsilon _{n}-A_{n}).  \label{a1}
\end{equation}
The linearization of (\ref{a1}) in a neighborhood of the fixed point (when $%
A_{n}=A^{\ast }$ and $\varepsilon _{n}=0$) is 
\begin{equation}
A_{n+1}=A^{\ast }+\left. \frac{\partial f}{\partial A_{n}}\right|
_{f.p.}(A_{n}-A^{\ast })+\left. \frac{\partial f}{\partial \varepsilon _{n}}%
\right| _{f.p.}\varepsilon _{n},  \label{a2}
\end{equation}
where 
%${\displaystyle \frac{\partial f}{\partial A_n} = \frac{d f}{d D_n} \frac{\partial D_n}{\partial A_n}}$,\ ${\displaystyle \frac{\partial f}{\partial \varepsilon_n} = \frac{d f}{d D_n} \frac{\partial D_n}{\partial \varepsilon_n}}$, and
$f.p.$ denotes evaluation at the fixed point. Since the controlled pacing
relation is $D_{n}=B+\varepsilon _{n}-A_{n}$, the derivatives in (\ref{a2})
are 
\begin{equation}
\left. \frac{\partial f}{\partial A_{n}}\right| _{f.p.}\ =\ \ -\left. \frac{%
\partial f}{\partial \varepsilon }\right| _{f.p.}\ =\ \ -\left. \frac{df}{%
dD_{n}}\right| _{f.p.}\ \equiv \ \mu .  \label{eq:deriv}
\end{equation}
We rewrite expressions (\ref{eq:epsilon}) and (\ref{a2}) in matrix form as 
\begin{equation}
\left( 
\begin{array}{c}
\delta _{n+1} \\ 
e_{n+1} \\ 
\alpha _{n+1}
\end{array}
\right) =\left( 
\begin{array}{ccc}
\mu & -\gamma \mu & 0 \\ 
-1 & 0 & 1 \\ 
1 & 0 & 0
\end{array}
\right) \left( 
\begin{array}{c}
\delta _{n} \\ 
e_{n} \\ 
\alpha _{n}
\end{array}
\right) ,  \label{eq:oldmatrix}
\end{equation}
where %\begin{equation}
$\delta _{n}=A_{n}-A^{\ast }$,\ %\qquad 
$e_{n}=\varepsilon _{n}/\gamma $, and %\qquad
$\alpha _{n}=\delta _{n-1}$. %\label{eq:param}
%\end{equation}
Control is successful when the eigenvalues of (\ref{eq:oldmatrix}) fall
within the unit circle in the complex plane. This occurs when $\mu $ and $%
\gamma $ lie within the region defined by the curves 
\begin{equation}
\mu =1,\quad \gamma =\frac{1+\mu }{2\mu },\quad \gamma =\frac{1-\mu -\sqrt{%
(1-\mu )^{2}+4}}{2\mu },  \label{eq:oldlines}
\end{equation}
as shown in Fig. 2a. Condition (\ref{eq:oldlines}) defines the domain of
control for the mapping model (\ref{eq:oldmap}). Note that the domain does 
{\em not} depend on the specific functional form of $f$, but only on the
value of its derivative $\mu $ evaluated at the fixed point. Alternans may
exist in the uncontrolled mapping model (\ref{eq:oldmap}) when $\mu \leq -1.$
The domain of control indicates the values of the gain $\gamma $ that should
be used to establish control of alternans for different values of the
Floquet multiplier $\mu $, which describes the stability of the map in the
absence of control. Note that the gain necessary to establish control in the
region where alternans exist in the absence of control is small ($0<\gamma <%
\sqrt{2}-1),$ in contradiction with the experimental observations \cite{dan}
where $\gamma $ can be as large as 4.
\begin{figure}
\includegraphics{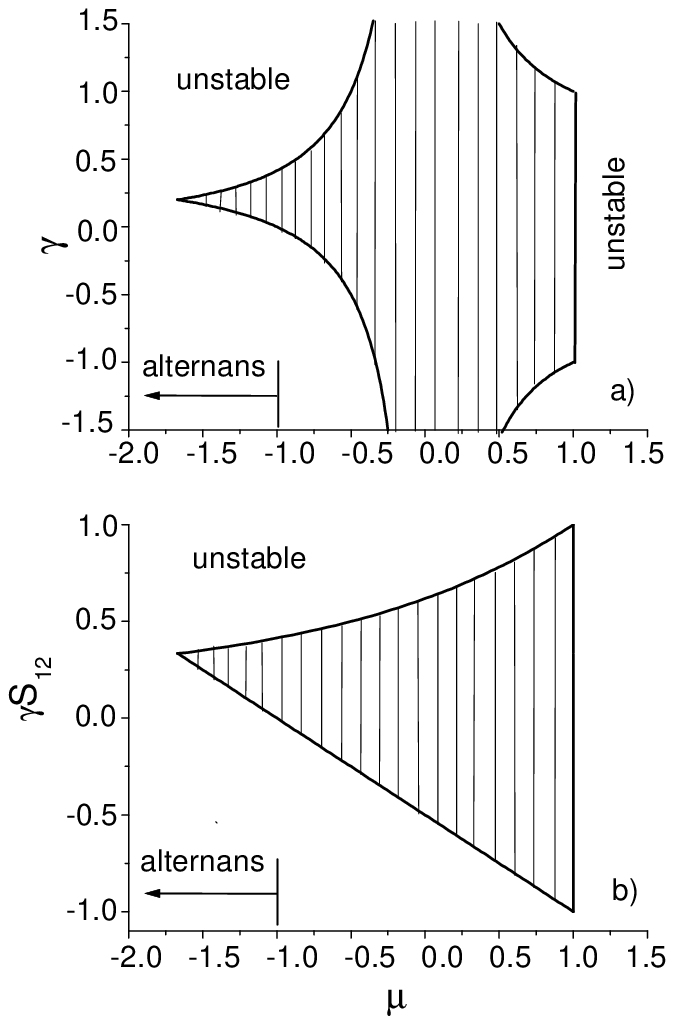}% Here is how to import EPS art
\caption{\label{fig:epsart} 
Domains of control of alternans (shaded region) for a) the
memoryless model (1) and b) the model with memory (2). Alternans may exist
in the absence of control when $\protect\mu \leq -1$.}
\end{figure}
Very different behavior is found for the mapping model (\ref{eq:map}) with
memory. In the presence of control, Eq. (\ref{eq:map}) becomes 
\begin{equation}
A_{n+1}=F(A_{n},D_{n})=F(A_{n},B+\varepsilon _{n}-A_{n}).  \label{eq:map1:1}
\end{equation}
Linearizing (\ref{eq:map1:1}) in a neighborhood of the fixed point, we find
that 
\begin{equation}
A_{n+1}=A^{\ast }+F^{\prime }\ (A_{n}-A^{\ast })+\left. \frac{\partial F}{%
\partial \varepsilon _{n}}\right| _{f.p.}\varepsilon _{n}.  \label{eq:taylor}
\end{equation}
where

\begin{equation}
F^{\prime }=\left. \frac{\partial F}{\partial A_{n}}\right| _{f.p.}-\left. 
\frac{\partial F}{\partial D_{n}}\right| _{f.p.}\equiv \mu , 
\qquad \left. 
\frac{\partial F}{\partial \varepsilon _{n}}\right| _{f.p.}=\left. \frac{%
\partial F}{\partial D_{n}}\right| =S_{12}. \end{equation}

As in Eq. (\ref{eq:deriv}), we denote $\mu $ as a full derivative of $F$,
evaluated at the fixed point. As seen from Eq. (\ref{eq:criterion}), the
condition for the existence of alternans in the absence of control is given
by $\mu \leq -1.$ Note that the value of $S_{12}$ does not predict the
existence of alternans; instead, it characterizes the immediate response of
the tissue to small perturbations from the fixed point, as was shown in Ref. 
\cite{lena1}. Hence, we expect that $S_{12}$ is more relevant for
determining the effects of control, whereas $\mu $ dictates the stability of
the steady-state response of the tissue.

We can rewrite expressions (\ref{eq:epsilon}) and (\ref{eq:taylor}) in
matrix form, using the same local variables as in (\ref{eq:oldmatrix}),

\begin{equation}
\left( 
\begin{array}{c}
\delta _{n+1} \\ 
e_{n+1} \\ 
\alpha _{n+1}
\end{array}
\right) =\left( 
\begin{array}{ccc}
\mu  & \gamma S_{12} & 0 \\ 
-1 & 0 & 1 \\ 
1 & 0 & 0
\end{array}
\right) \left( 
\begin{array}{c}
\delta _{n} \\ 
e_{n} \\ 
\alpha _{n}
\end{array}
\right) .  \label{eq:matrix}
\end{equation}
The only difference between the matrices describing the controlled dynamics
for the maps with and without memory is the term containing the control gain 
$\gamma $. In the\ memoryless mapping model (\ref{eq:oldmap}), the magnitude
of perturbation sensitivity is proportional to the full derivative $\mu $,
whereas it is proportional to the slope $S_{12}$ of the S1-S2 RC in the
mapping model with memory (\ref{eq:map}). When there is no memory, $%
S_{12}=-\mu $, whereas they can differ substantially when memory is present.

By analyzing (\ref{eq:matrix}), we find that control is successful when $\mu 
$ and $\gamma S_{12}$ lie within the region defined by the curves 
\begin{eqnarray}
\mu &=&1,\qquad \gamma S_{12}=-(\mu +1)/2,\qquad  \nonumber \\
\gamma S_{12} &=&-\left( 1-\mu -\sqrt{(1-\mu )^{2}+4}\right) /2.
\label{eq:lines1}
\end{eqnarray}
Similar to the previous model, the domain of control (\ref{eq:lines1}) does
not depend on the specific form of the function $F$. We only need to know
the values of the full derivative $\mu $ and the slope of the S1-S2 RC $%
S_{12}$ (evaluated at the fixed point) to determine whether it is possible
to establish control. Figure 2b depicts the domain of control for the
one-dimensional mapping model with memory (\ref{eq:map}) according to
conditions (\ref{eq:lines1}). Alternans may exist in the uncontrolled system
when $\mu \leq -1.$ It can be seen from Fig. 2b that the value of $\gamma
S_{12}$ necessary to establish control for this region is relatively small ($%
0<\gamma S_{12}<\sqrt{2}-1$). However, the actual control gain $\gamma $ can
be larger or smaller than this range depending on the value of $S_{12}$. Our
predictions are consistent with our earlier expectations that $S_{12}$
influences the effects of control since it quantifies the response of the
tissue (in equilibrium) to a sudden perturbation, which the controller
attempts to cancel.

Comparing our predictions to experiments is not possible at this time
because no experiments have measured the slope of the S1-S2 RC {\em %
evaluated at the fixed point} \cite{note}. Most of the experiments (see, for
instance, Refs. \cite{gilmour2,mary}) used S1-S2 pacing protocol in which
the ``S1'' interval was either fixed or set to only a few different values.
Instead, to make a direct comparison of the domain of control obtained using
mapping model (\ref{eq:map}) to the one obtained experimentally in Ref. \cite
{dan}, we need to know the slope of the S1-S2 RC $S_{12}$ at each point on
the dynamic RC, {\it i.e.}, for different S1. The value $S_{12}$ can be
greater or less than one, depending on the specific form of $F$ or the
specific type of tissue. For example, $S_{12}>1$ for the function $F$ used
to generate the plot shown in Fig. 1b. Hence, the control gain $\gamma $
must be less than ($\sqrt{2}-1)$ to be in the region where control is
effective. However, preliminary, experiments with bullfrog cardiac muscle
indicate that $S_{12}$ can be relatively small (less than 0.4 and as small
as 0.05) at the onset of alternans \cite{soma} and thus the control gain
could be large.

Thus, the experimentally measured domain of control may be consistent with
the predictions of the controlled map with memory (\ref{eq:map}) if $S_{12}$
is truly less than one in bullfrog. The map without memory (\ref{eq:oldmap})
does not agree with experiments as demonstrated by Fig. 2a and as noted
previously in Ref. \cite{dan}. Our analysis is the first to suggest that
memory effects may substantially enlarge the domain of control, regulated by 
$S_{12}.$ Future experiments are needed to clarify the effects of memory on
control.

A limitation of our analysis is that the extent of cardiac memory is only
one previous beat in the model (\ref{eq:map}), so that the memory is
short-term. However, some studies indicate that long-term memory effects are
present in real cardiac tissue \cite{mary} that might be described using a
long-term memory mapping model \cite{fox,chialvo}. In this model, an
additional memory variable is introduced, which is assumed to accumulate
during the APD and dissipate during the DI. Analysis of this model \cite{fox}
confirms that the slope of the dynamic RC does not predict the transition to
alternans, but it was shown in Ref. \cite{dan} that a specific form of such
a memory model cannot explain the experiments on control of alternans. We
believe that a generalization of the mapping model (\ref{eq:map}) that
includes long-term memory as in \cite{fox,chialvo} may be a fruitful
direction for future investigations.

We gratefully acknowledge the support of the National Science Foundation
under Grant PHY-9982860 and DMS-9983320 (M.M.R).


\begin{references}
\bibitem{note}  We note that a pacing protocol has been suggested in [18]
for easily determining $S_{dyn}$, $S_{12}$, and $F^{\prime }$ (evaluated at
the fixed points) in experiments.
\end{references}
\end{document}